\documentclass[onecollarge,natbib]{svjour2}
\bibpunct{[}{]}{;}{n}{}{,} 
\smartqed  
\usepackage{graphicx}
\journalname{Archive of Applied Mechanics}
\begin{document}

\title{Updates on the Studies of $N^*$ Structure with CLAS and the Prospects with CLAS12
}


\author{V.I. Mokeev for the CLAS Collaboration}


\institute{V.I. Mokeev \at
              Thomas Jefferson National Accelerator Facility, Newport News, VA 23606 \\
              Tel.: +1-757-269-6990\\
              Fax: +1-757-269-5800\\
              \email{mokeev@jlab.org}           
           \and
}

\date{Received: date / Accepted: date}

\maketitle

\begin{abstract}
The recent results on $\gamma_vpN^*$ electrocouplings from analyses of the data on exclusive meson 
electroproduction off protons measured with the CLAS detector at Jefferson Lab are presented. The impact 
of these results on the exploration of the excited nucleon state structure and non-perturbative strong 
interaction dynamics behind its formation is outlined. The future extension of these studies in the 
experiments with the CLAS12 detector in the upgraded Hall-B at JLab  will provide for the first time 
$\gamma_vpN^*$ electrocouplings of all prominent resonances at the still unexplored distance scales 
that correspond to extremely low (0.05~GeV$^2 < Q^2 <$ 0.5~GeV$^2$) and to the highest photon virtualities 
(5.0~GeV$^2 < Q^2 <$ 12.0~GeV$^2$) ever achieved in the exclusive electroproduction measurements. 
The expected results will address the most important open problems of the Standard Model: on the nature 
of more than 98\% of hadron mass, quark-gluon confinement and the emergence of the excited nucleon state 
structure from the QCD Lagrangian, as well as allowing a search for the new states of hadron matter 
predicted from the first principles of QCD, the so-called hybrid baryons.   
\end{abstract}

\section{Introduction}
\label{intro}

The studies of helicity amplitudes that describe the transitions between the initial real/virtual 
photon - ground state proton and the final $N^*/\Delta^*$ states, the so-called $\gamma_{r,v}pN^*$ 
photo-/electrocouplings, represent an important and absolutely needed effort in the exploration 
of non-perturbative strong interaction dynamics behind the generation of the ground and excited nucleon 
states from quarks and gluons. These studies, carried out in a wide range of photon virtualities $Q^2$ 
and over all prominent excited nucleon states, are the only source of information on different 
manifestations of the non-perturbative strong interaction in the generation of excited nucleons of 
different quantum numbers~\cite{Br15,Bu15,Mo16}. Furthermore, the recent studies of nucleon structure 
within the framework of the Dyson-Schwinger Equations of QCD (DSEQCD)~\cite{Seg15,Seg14,Cr15,Cr16} 
conclusively demonstrated the critical importance of combined analysis of the data on elastic and 
transition $p \to N^*$ form factors in order to provide credible access to the momentum dependence of 
the running dynamical mass of dressed quarks. This fundamental ingredient of the non-perturbative strong 
interaction elucidates the generation of more than 98\% of hadron mass and the emergence of quark-gluon 
confinement. It makes the studies of the elastic  and transition  $p \to N^*$ form factors, which are 
directly related to $\gamma_{r,v}pN^*$ photo-/electrocouplings, one of the central focuses of contemporary 
hadron physics. 

In this proceedings we outline the current status in the studies of the $\gamma_vpN^*$ electrocouplings
from the data on exclusive meson electroproduction off protons measured with the CLAS detector at JLab 
\cite{Bu15,Mo16,Bu12,Az13,Bu16,Mo16a}, as well as the insight to the structure of excited nucleon states 
offered by these results. We discuss also the prospects for the future extension of the resonance 
electrocoupling studies in the experiments foreseen with the CLAS12 detector. The results expected from 
these experiments will address the key open problems of the Standard Model on the nature of more than 
98\% of the ground and excited nucleon masses, the emergence of quark-gluon confinement and the nucleon spectrum and structure from QCD~\cite{Cr14}. Furthermore, they 
will allow us to search for the new states of baryon matter predicted in the Lattice QCD studies of the 
$N^*$ spectrum~\cite{Du12}, the so-called the hybrid baryons~\cite{Bu16,La16}.

\section{Evaluation of $\gamma_vpN^*$ electrocouplings from the CLAS data on exclusive meson 
electroproduction off protons}
\label{sec1}

The CLAS detector has contributed the lion's share of the world data on all essential exclusive 
meson electroproduction channels in the resonance excitation region including $\pi N$, $\eta p$, $KY$, 
and $\pi^+\pi^-p$ electroproduction off protons with nearly complete coverage of the final hadron phase 
space~\cite{Bu12}. The observables measured with the CLAS detector are stored in the CLAS Physics Data 
Base~\cite{db15}. The available observables in the resonance excitation region are listed in 
Table~\ref{tab-1}. In the near future the data in the resonance region will be extended by new results 
on $\pi^+ n$, $\pi^0 p$, and  $\pi^+\pi^-p$ exclusive electroproduction at $W < 1.9$~GeV and 
0.3~GeV$^2$ $< Q^2 <$ 1.0~GeV$^2$. The new data on $\pi^+\pi^-p$ electroproduction off protons at 
$W < 2.0$~GeV will become available in the $Q^2$ range from 2.0~GeV$^2$ to 5.0~GeV$^2$. 

So far, most of the results on the $\gamma_vpN^*$ electrocouplings have been extracted from independent 
analyses of $\pi^+n$, $\pi^0p$, and $\pi^+\pi^-p$ exclusive electroproduction data off the proton. A 
total of nearly 160,000 data points (d.p.) on unpolarized differential cross sections, longitudinally 
polarized beam asymmetries, and longitudinal target and beam-target asymmetries for $\pi N$ electroproduction 
off protons were obtained with the CLAS detector at $W < 2.0$~GeV and 0.2~GeV$^2$ $< Q^2 < 6.0$~GeV$^2$. The 
data have been analyzed within the framework of two conceptually different approaches: a unitary isobar model 
(UIM) and dispersion relations (DR)~\cite{Az09,Park15}. The UIM describes the $\pi N$ electroproduction 
amplitudes as a superposition of $N^*$ electroexcitations in the $s$-channel, non-resonant Born terms, and 
$\rho$- and $\omega$- $t$-channel contributions. The latter are reggeized, which allows for a better 
description of the data in the second- and third-resonance regions. The final-state interactions are treated 
as $\pi N$ rescattering in the K-matrix approximation~\cite{Az09}. In the DR approach, dispersion relations 
relate the real to the imaginary parts of the invariant amplitudes that describe the $\pi N$ electroproduction. 
Both approaches provide a good and consistent description of the $\pi N$ data in the range of $W < 1.7$~GeV 
and $Q^2 < 5.0$~GeV$^2$, resulting in $\chi^2/d.p. < 2.9$. 
  
The $\pi^+\pi^- p$ electroproduction data from CLAS~\cite{Ri03,Fe09} provide information for the first time
on nine independent single-differential and fully-integrated cross sections binned in $W$ and $Q^2$ in the 
mass range $W < 2.0$~GeV and at photon virtualities of 0.25~GeV$^2 < Q^2 < 1.5$~GeV$^2$. The analysis of the 
data have allowed us to develop the JM reaction model~\cite{Mo16,Mo12,Mo09} with the goal of extracting 
resonance electrocouplings, as well as $\pi\Delta$ and $\rho p$ hadronic decay widths. This model incorporates 
all relevant reaction mechanisms in the $\pi^+\pi^-p$ final-state channel that contribute significantly to the 
measured electroproduction cross sections off protons in the resonance region, including the 
$\pi^-\Delta^{++}$, $\pi^+\Delta^0$, $\rho^0 p$, $\pi^+N(1520)\frac{3}{2}^-$, $\pi^+N(1685)\frac{5}{2}^+$, 
and $\pi^-\Delta(1620)\frac{3}{2}^+$ meson-baryon channels, as well as the direct production of the 
$\pi^+\pi^-p$ final state without formation of intermediate unstable hadrons. In collaboration with JPAC 
\cite{Szcz15} a special approach has been developed allowing us to remove the contributions from the  
$s$-channel resonances to the reggeized $t$-channel non-resonant terms in the $\pi^-\Delta^{++}$, 
$\pi^+\Delta^0$, $\rho^0 p$ electroproduction amplitudes. The contributions from well established $N^*$ 
states in the mass range up to 2.0~GeV were included into the amplitudes of the $\pi\Delta$ and $\rho p$ 
meson-baryon channels by employing a unitarized version of the Breit-Wigner ansatz~\cite{Mo12}. The JM model 
provides a good description of the $\pi^+\pi^- p$ differential cross sections at $W < 1.8$~GeV and 
0.2~GeV$^2 < Q^2 < 1.5$~GeV$^2$ with $\chi^2/d.p. < 3.0$. The quality for the description of the CLAS 
data suggests the unambiguous and credible separation between the resonant/non-resonant contributions achieved 
fitting the CLAS data~\cite{Mo16}. The credible isolation of the resonant contributions makes it possible to 
determine the resonance electrocouplings, along with the $\pi \Delta$, and $\rho N$ decay widths from the 
resonant contributions employing for their description the amplitudes of the unitarized Breit-Wigner ansatz
\cite{Mo12} that fully accounts for the unitarity restrictions on the resonant amplitudes. 

\begin{table}[htb!]
\begin{center}
\caption{\label{tab-1} Observables for exclusive meson electroproduction off protons that have been 
measured with the CLAS detector in the resonance excitation region and stored in the CLAS Physics Data Base 
\cite{db15}: CM-angular distributions for the final mesons ($\frac{d\sigma}{d\Omega}$); beam, target, and 
beam-target asymmetries ($A_{LT'}$, $A_t$, $A_{et}$); and recoil hyperon polarizations ($P'$, $P^0$).}
\begin{tabular}{|c|c|c|c|}
\hline
Hadronic       &  $W$-range    & $Q^2$-range     & Measured observables \\
final state    &  GeV          & GeV$^2$         &     \\ 
\hline
$\pi^+ n$      &  1.10-1.38     & 0.16-0.36      & $\frac{d\sigma}{d\Omega}$ \\
               &  1.10-1.55     & 0.30-0.60      & $\frac{d\sigma}{d\Omega}$ \\
               &  1.10-1.70     & 1.70-4.50      & $\frac{d\sigma}{d\Omega}$, $A_{LT'}$ \\
               &  1.60-2.00     & 1.80-4.50      &  $\frac{d\sigma}{d\Omega}$    \\
\hline	       
$\pi^0 p$      &  1.10-1.38     & 0.16-0.36      & $\frac{d\sigma}{d\Omega}$ \\
               &  1.10-1.68     & 0.40-1.15      & $\frac{d\sigma}{d\Omega}$, $A_{LT'}$, $A_t$, $A_{et}$ \\
               &  1.10-1.39     & 3.00-6.00      & $\frac{d\sigma}{d\Omega}$  \\
\hline     
$\eta p$       &  1.50-2.30     & 0.20-3.10      & $\frac{d\sigma}{d\Omega}$ \\
\hline     
$K^+\Lambda$   &  1.62-2.60     & 1.40-3.90      & $\frac{d\sigma}{d\Omega}$ \\
               &  1.62-2.60     & 0.70-5.40      & $P'$, $P^0$ \\
\hline     
$K^+\Sigma^0$   &  1.62-2.60     & 1.40-3.90     & $\frac{d\sigma}{d\Omega}$ \\
                &  1.62-2.60     & 0.70-5.40     & $P'$ \\
\hline     
$\pi^+\pi^-p$   &  1.30-1.60     & 0.20-0.60      & Nine single-differential \\
                &  1.40-2.10     & 0.50-1.50      & cross sections \\
\hline
\end{tabular}
\end{center}
\end{table}

Electrocouplings of nucleon resonances and their $KY$ hadronic decay widths can also be determined from 
analyses of the CLAS data on exclusive $KY$ electroproduction off protons~\cite{Ca16}. Future analyses 
of the large body of these data (Table~\ref{tab-1}) will improve the knowledge on electrocouplings and $KY$ hadronic decay widths of 
high-lying $N^*$ states with masses above 1.6~GeV. The decay to the $\pi N$ final state for many of these 
resonances are suppressed. Currently, the preliminary results on the electrocouplings of these states are 
available from the studies of $\pi^+\pi^-p$ electroproduction off protons only \cite{Mo16a,Mo16b}. Consistent results on 
$\gamma_vpN^*$ electrocouplings of the aforementioned resonances obtained in independent analyses of $KY$ 
electroproduction will validate credible extraction of these fundamental quantities. The development of the 
analysis tools for extraction of the resonance parameters from $KY$ electroproduction data measured with CLAS 
is urgently needed.     

\section{Selected results on resonance electrocouplings and their impact on the insight into 
$N^*$ structure} 
\label{results}

Resonance electrocouplings have been obtained from various CLAS data in the exclusive channels: $\pi^+n$ 
and $\pi^0p$ at $Q^2 < 5.0$~GeV$^2$ in the mass range up to 1.7~GeV, $\eta p$ at $Q^2 < 4.0$~GeV$^2$ in 
the mass range up to 1.6~GeV, and $\pi^+\pi^-p$ at $Q^2 < 1.5$~GeV$^2$ in the mass range up to 1.8~GeV 
\cite{Mo16,Bu12,Az09,Park15,Mo12,Mo16b}. The studies of the $N(1440)1/2^+$ and $N(1520)3/2^-$ resonances with the CLAS 
detector~\cite{Mo16,Az09,Mo12} have provided the dominant part of the information available worldwide on 
their electrocouplings in a wide range of photon virtualities 0.25~GeV$^2 < Q^2 < 5.0$~GeV$^2$. 
Currently the $N(1440)1/2^+$ and $N(1520)3/2^-$ states, together with the $\Delta(1232)3/2^+$ and 
$N(1535)1/2^-$ resonances~\cite{Bu12}, represent the most explored excited nucleon states. Furthermore, 
results on the $\gamma_vpN^*$ electrocouplings for the high-lying $N(1675)5/2^-$, $N(1680)5/2^+$, and 
$N(1710)1/2^+$ resonances have been determined for the first time from the CLAS $\pi N$ data at 
1.5~GeV$^2 < Q^2 < 4.5$~GeV$^2$~\cite{Park15}. The first results on the electrocouplings of 
$\Delta(1620)1/2^-$ resonance that preferentially decays to the $N\pi\pi$ final states have recently 
become available from analysis of the CLAS data on $\pi^+\pi^-p$ electroproduction off protons. The up to 
date numerical results on electrocouplings of most nucleon resonances in the mass range up to 1.8~GeV 
available from the exclusive electroproduction data from CLAS and elsewhere are maintained on our
web page~\cite{Mo16b}.   

\begin{figure}[htb!]
\centering
\includegraphics[width=4.6cm,clip]{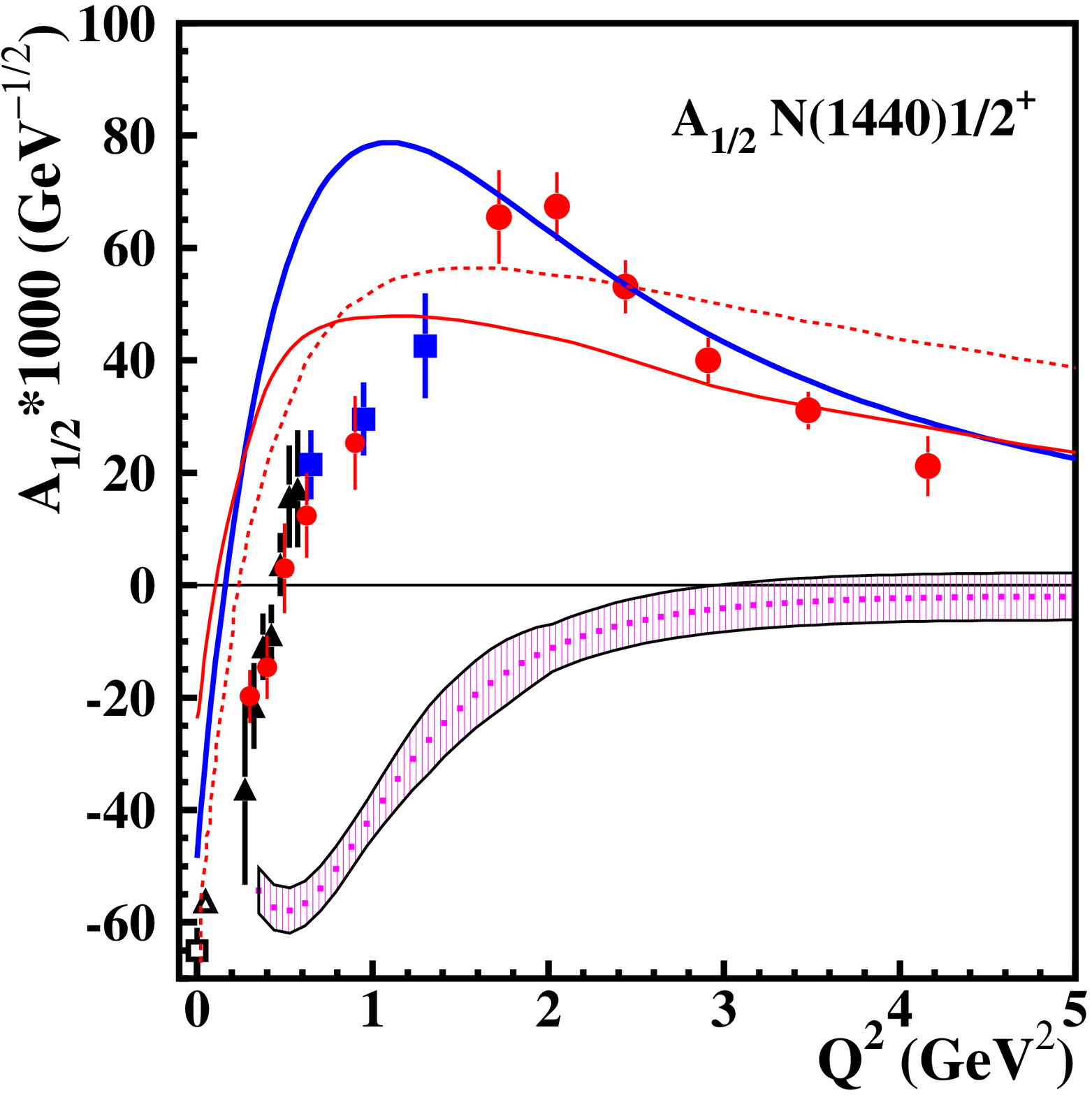}
\includegraphics[width=4.6cm,clip]{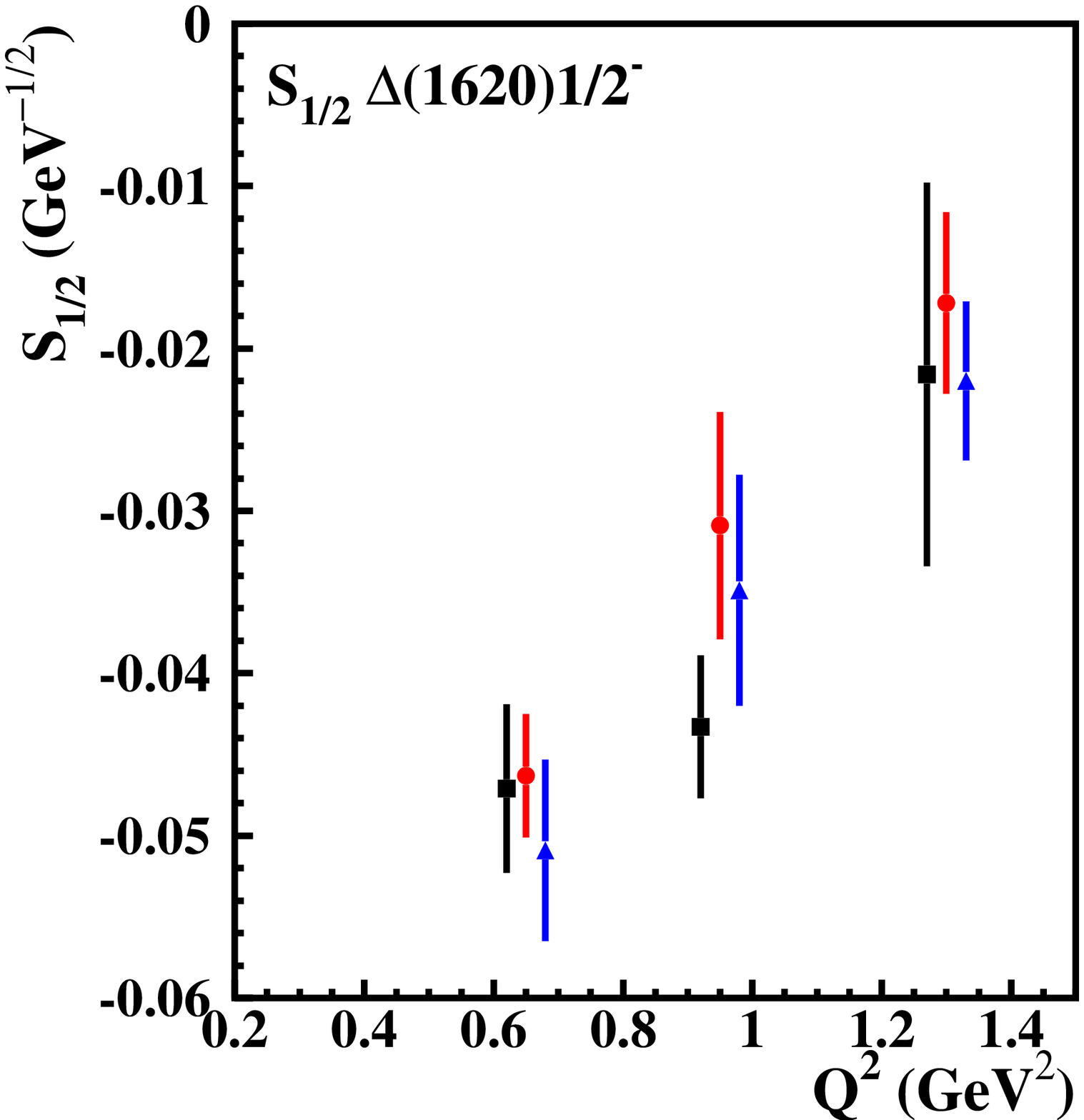}
\includegraphics[width=4.6cm,clip]{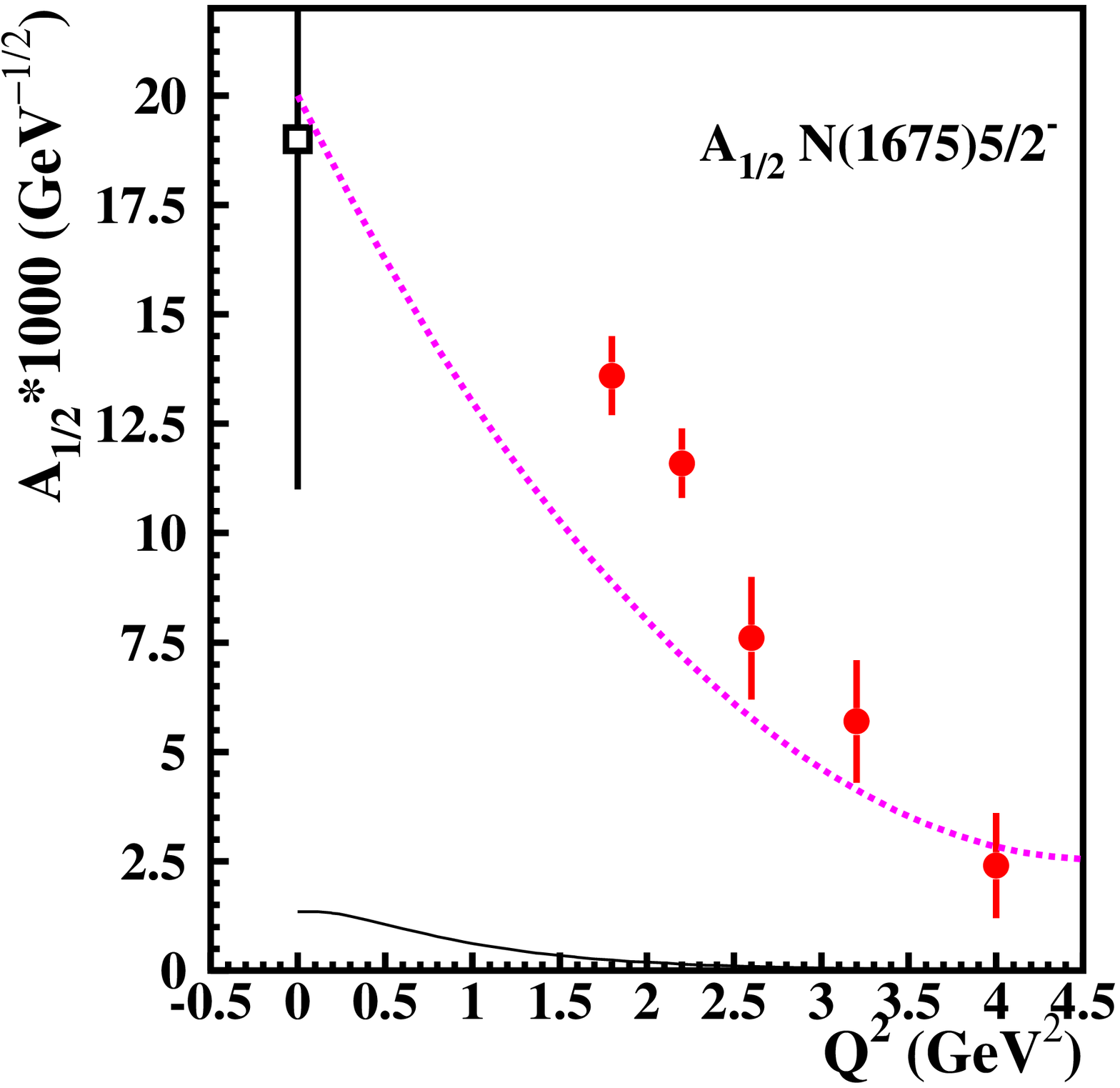} 
\caption{(Color Online) $A_{1/2}$ $\gamma_vpN^*$ electrocouplings of the $N(1440)1/2^+$ (left), 
$N(1675)5/2^-$ (right), and  $S_{1/2}$ $\gamma_vpN^*$ electrocouplings of the $\Delta(1620)1/2^-$ 
(center) resonances from analyses of the CLAS electroproduction data off protons in the $\pi N$ - 
\cite{Az09,Park15} (red circles in the left and right panels) and $\pi^+\pi^-p$ channels~\cite{Mo12} 
(black triangles in the left panel), with new results from the $\pi^+\pi^-p$ channel~\cite{Mo16} 
(blue squares in the left panel). The central panel shows the $\Delta(1620)1/2^-$ electrocouplings 
obtained from analyses of $\pi^+\pi^-p$ electroproduction data off protons \cite{Mo16} carried out 
independently in three intervals of $W$: 1.51~GeV $\to$ 1.61~GeV (black squares), 1.56~GeV $\to$ 
1.66~GeV (red circles), and 1.61~GeV $\to$ 1.71~GeV (blue triangles). Photocouplings are taken from the
RPP~\cite{rpp} (open squares) and the CLAS data analysis~\cite{Dug09} of $\pi N$ photoproduction (open 
triangles). The electrocoupling results in the left panel are shown in comparison with the DSEQCD - 
\cite{Seg15} (blue thick solid) and constituent quark model calculations~\cite{Az12} (thin red solid), 
\cite{Ob14} (thin red dashed). The meson-baryon cloud contributions, determined as described in 
Section~\ref{results}, are shown by the magenta area. For the case of the $N(1675)5/2^-$ resonance (right), 
the absolute values of the meson-baryon cloud amplitudes at the resonance poles taken from Argonne-Osaka 
coupled channel analysis~\cite{Lee08} are shown by dashed magenta line together with the estimate for the
quark core contribution from a quark model~\cite{Sa12} (black solid line).}
\label{p11d13d15}
\end{figure}

\begin{figure}[htb!]
\centering
\includegraphics[width=4.6cm,clip]{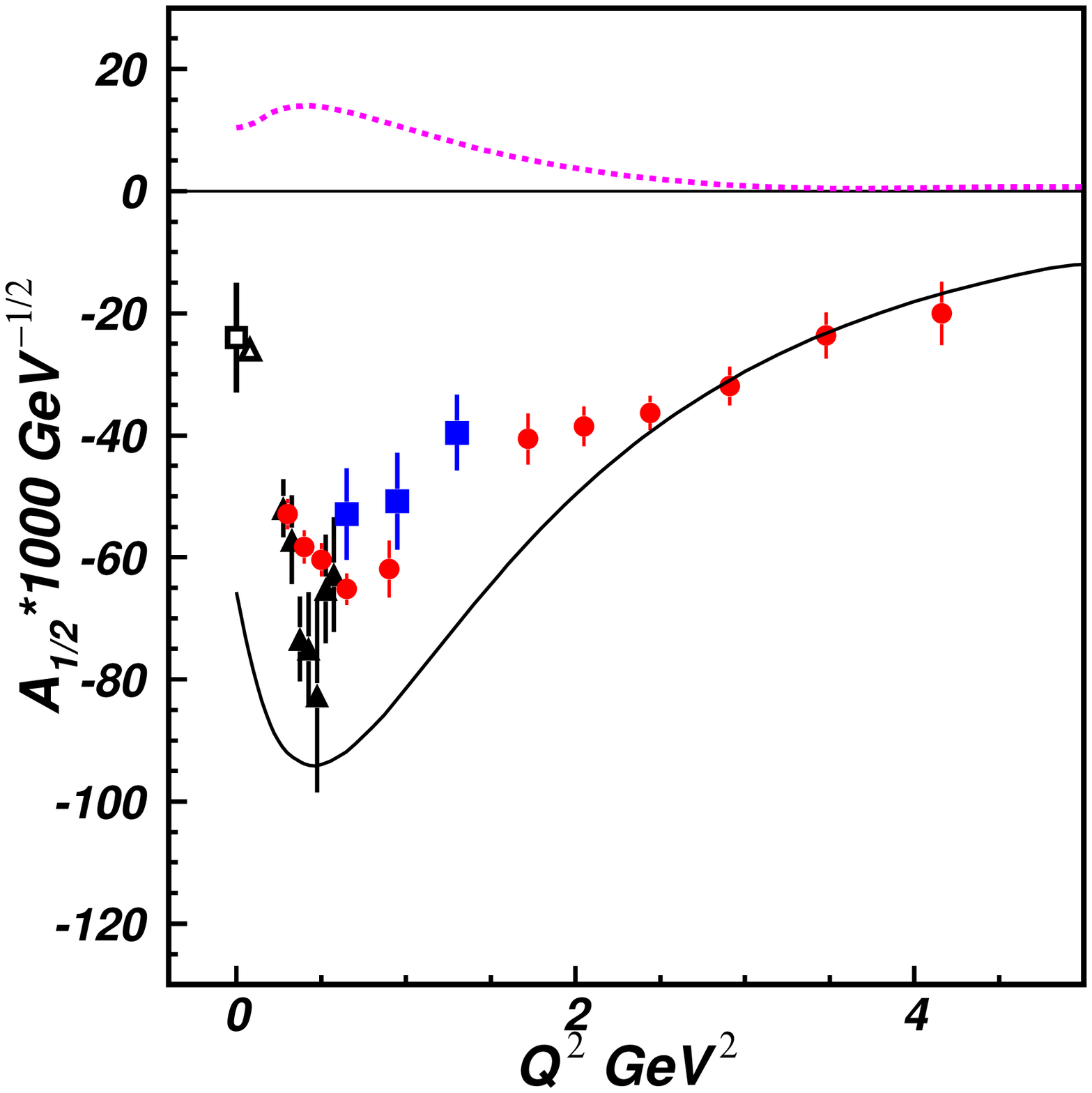}
\includegraphics[width=6.3cm,clip]{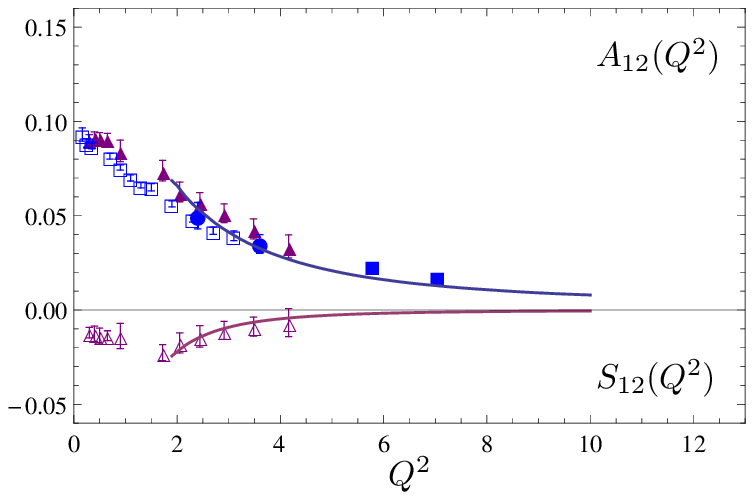} 
\caption{(Color Online) $A_{1/2}$ photo-/electrocouplings of the $N(1520)3/2^-$ (left) and $N(1535)1/2^-$ 
(right) resonances from analyses of the CLAS electroproduction data off protons~\cite{Mo16,Az09,Mo12,Den07}. 
The symbol meaning in the left panel is the same as in Fig.~\ref{p11d13d15} (left). The $N(1535)1/2^-$ 
electrocouplings in the right panel from $\pi N$~\cite{Az09} and $\eta N$~\cite{Den07} electroproduction 
are shown by the triangles and rectangles, respectively. The analysis~\cite{Den07} of the $\eta N$ data 
was carried out assuming $S_{1/2}$=0. The electrocoupling results for the $N(1520)3/2^-$ are shown in 
comparison with the model estimates for the quark core~\cite{Sa12} (black solid). The meson-baryon cloud 
contributions taken from the Argonne-Osaka coupled channel analysis~\cite{Lee08} are shown by the magenta 
dashed line. In case of the $N(1535)1/2^-$ resonance (right) the data are compared with the results of the 
LCSR model~\cite{Br15} with the quark DA normalization parameters from the LQCD results obtained starting 
from the QCD Lagrangian~\cite{Br14}.}
\label{d13s11}
\end{figure}

Consistent results for the $\gamma_vpN^*$ electrocouplings of the $N(1440)1/2^+$ and $N(1520)3/2^-$
resonances, which have been determined in independent analyses of the dominant meson electroproduction 
channels, $\pi N$ and $\pi^+\pi^-p$ shown in Fig.~\ref{p11d13d15} (left) and in Fig.~\ref{d13s11} (left), 
demonstrate that the extraction of these fundamental quantities is reliable, since a good data description 
is achieved in the major electroproduction channels having quite different background contributions. We 
have developed special procedures to test the reliability of the resonance $\gamma_vpN^*$ electrocouplings 
extracted from the charged double pion electroproduction data only. In this case, we carried out the 
extraction of the resonance parameters independently, fitting the CLAS $\pi^+\pi^-p$ electroproduction data
\cite{Ri03} in overlapping intervals of $W$. The non-resonant amplitudes in each of the presented in Fig~\ref{p11d13d15} (center) $W$-intervals are different, while the resonance parameters should remain the same as they are determined 
from the data fit in different $W$-intervals. The electrocouplings of the $\Delta(1620)1/2^-$ state 
determined in this procedure are shown in the center of Fig.~\ref{p11d13d15}. The consistent results on 
these electrocouplings from the independent analyses of different $W$-intervals strongly support their 
reliable extraction. The tests described above demonstrated the capability of the models outlined in Section~\ref{sec1}  to provide reliable information on the $\gamma_vpN^*$ resonance electrocouplings from 
independent analyses of the data on exclusive $\pi N$ and $\pi^+\pi^-p$ electroproduction.
  
Experimental results on $\gamma_vpN^*$ electrocouplings offer insight into the non-perturbative strong 
interaction dynamics behind the emergence of the $N^*$ structure from the QCD Lagrangian. Two 
conceptually different approaches have been developed that allow us to relate the information on the
$Q^2$ evolution of the resonance electrocouplings to the first principles of QCD.

Due to the rapid progress in the field of DSEQCD studies of excited nucleon states
\cite{Seg15,Seg14,Cr15,Cr16a}, the first evaluations of the transition $p \to \Delta(1232)3/2^+$ 
form factors and the $N(1440)1/2^+$ resonance electrocouplings starting from the QCD Lagrangian have 
recently become available. The $A_{1/2}$ electrocouplings of the $N(1440)1/2^+$ resonance computed in 
\cite{Seg15} are shown in Fig.~\ref{p11d13d15} (left) by the solid blue lines. The evaluation~\cite{Seg15} is applicable at photon virtualities $Q^2 > 2.0$~GeV$^2$ where the contributions of the inner quark core to the resonance 
electrocouplings are much larger than from the external meson baryon cloud. In this range of photon 
virtualities, the evaluation~\cite{Seg15} offers a good description of the experimental results on the 
transition $p \to \Delta(1232)3/2^+$ form factors and the $N(1440)1/2^+$ resonance electrocouplings achieved 
with a momentum dependence of the dressed quark mass that is {\it exactly the same} as the one employed in 
the previous evaluations of the elastic electromagnetic nucleon form factors~\cite{Seg14}. This success 
strongly supports: a) the relevance of dynamical dressed quarks, with properties predicted by the DSEQCD 
approach~\cite{Cr14}, as constituents of the quark core in the structure of the ground and excited nucleon 
states; and b) the capability of the DSEQCD approach~\cite{Seg15,Seg14} to map out the dressed quark mass 
function from the experimental results on the $Q^2$-evolution of the nucleon elastic - and $p \to N^*$ 
electromagnetic transition form factors or rather $\gamma_vpN^*$ electrocouplings. 

The model~\cite{Br15} employs the light cone some rules (LCSR) in order to relate the quark 
distribution amplitudes (DA) of the excited nucleon states to the $Q^2$ evolution of their 
electrocouplings. Analysis of the $N(1535)1/2^-$ $\gamma_vpN^*$ electrocouplings from CLAS within the 
framework of this model has provided access to the quark DA's of excited nucleons for the first time. 
The model~\cite{Br15} describes successfully the CLAS results shown in Fig.~\ref{d13s11} (right) at 
$Q^2 > 2.0$~GeV$^2$ where LCSR's are applicable, with the values of the normalization parameters for 
the leading twist $N(1535)1/2^-$ quark DA obtained from the QCD Lagrangian within the framework of 
lattice QCD~\cite{Br14}. The results on the $N(1535)1/2^-$ electrocouplings in the much broader range of 
photon virtualities up to 12~GeV$^2$ expected from the future $N^*$ studies with the CLAS12 detector in 
Hall~B~\cite{Ca16,Go16,Go16a} will considerably improve knowledge of the $N(1535)1/2^-$ quark DA, allowing 
us to determine from electrocoupling data not only the normalization parameters, but also most of the 
shape parameters of the quark DA's. These studies will be extended by analyses of other excited nucleon 
states. Confronting the quark DA's of excited nucleon states determined from the experimental results on 
$\gamma_vpN^*$ electrocouplings to the lattice QCD expectations, offers an alternative way with respect 
to DSEQCD approaches to study the emergence of the resonance structure from the first principles of the 
QCD. 

The quark DA's of excited nucleon states are also expected from the future DSEQCD studies~\cite{Cr16}. 
Consistency between the expectations for the $N^*$ quark DA parameters obtained within the framework of 
LQCD/DSEQCD  and their values from the fit to the data on resonance electrocouplings within the framework 
of LCSR will validate credible access to the fundamental ingredients of the non-perturbative strong 
interaction behind the emergence of the $N^*$ structure from the QCD Lagrangian.

Analysis of the CLAS results on the $\gamma_vpN^*$ electrocouplings over most excited nucleon states in 
the mass range up to 1.8~GeV has revealed the $N^*$ structure for $Q^2 < 5.0$~GeV$^2$ as a complex 
interplay between an inner core of three dressed quarks and an external meson-baryon cloud. The two 
extended quark models~\cite{Az12,Ob14}, that account for the meson-baryon cloud and the quark core 
contributions combined, provided a better description of the $N(1440)1/2^+$ electrocouplings shown in 
Fig.~\ref{p11d13d15} (left) at $Q^2 < 2.0$~GeV$^2$, demonstrating the increasing importance of the 
meson-baryon contributions as $Q^2$ decreases. The credible DSEQCD evaluation of the quark core 
contributions to the electrocouplings of the $N(1440)1/2^+$ state has allowed us to derive the meson-baryon 
cloud contributions to this resonance as the difference between the experimental data on resonance 
electrocouplings and the quark core electroexcitation amplitudes from DSEQCD~\cite{Seg15} shown by the 
blue line in Fig~\ref{p11d13d15} (left). The obtained meson-baryon cloud contributions are presented in 
Fig.~\ref{p11d13d15} (left) by magenta dashed area. The relative contributions of the quark core and the
meson-baryon cloud depend strongly on the quantum numbers of the excited nucleon state. The quark core 
becomes the dominant contributor to the $A_{1/2}$ electrocouplings of the $N(1440)1/2^-$ and $N(1520)3/2^-$ 
resonances at $Q^2 > 2.0$~GeV$^2$, as can be seen in Fig.~\ref{p11d13d15} (left) and Fig.~\ref{d13s11} 
(left), respectively. The results on these state electrocouplings offer almost direct access to the dressed 
quark contributions for $Q^2 > 2.0$~GeV$^2$. Instead, electrocouplings of the $N(1675)5/2^-$ state, shown in 
Fig~\ref{p11d13d15} (right), are dominated by meson-baryon cloud, allowing us to explore this component from 
the electrocoupling data. The relative contributions of the meson-baryon cloud to the electrocouplings of all 
resonances studied with the CLAS decrease with $Q^2$ in gradual transition towards quark core dominance at 
photon virtualities above 5.0~GeV$^2$. 
 
\begin{figure}[htb!]
\centering
\includegraphics[width=8.0cm,clip]{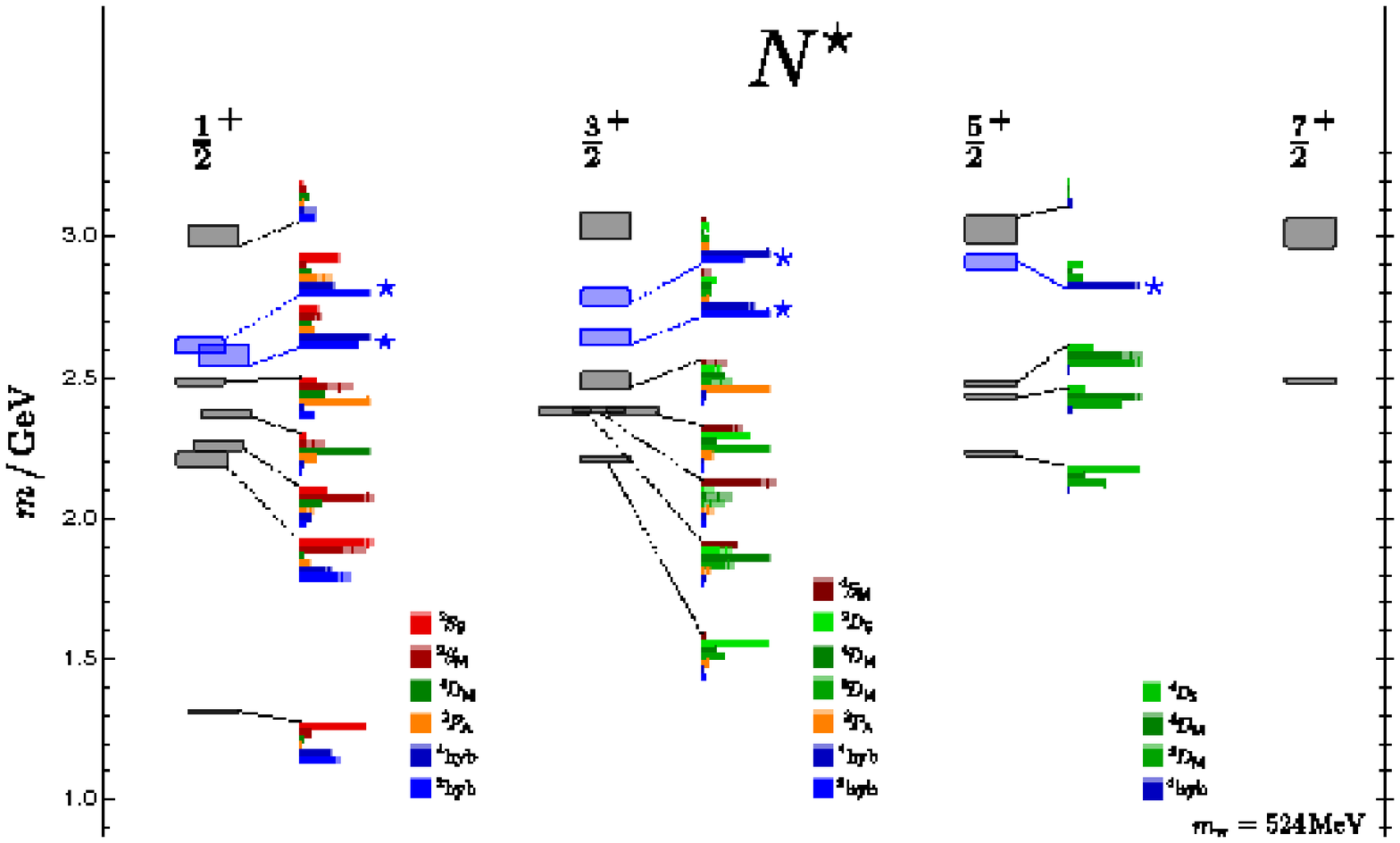}
\includegraphics[width=6.8cm,clip]{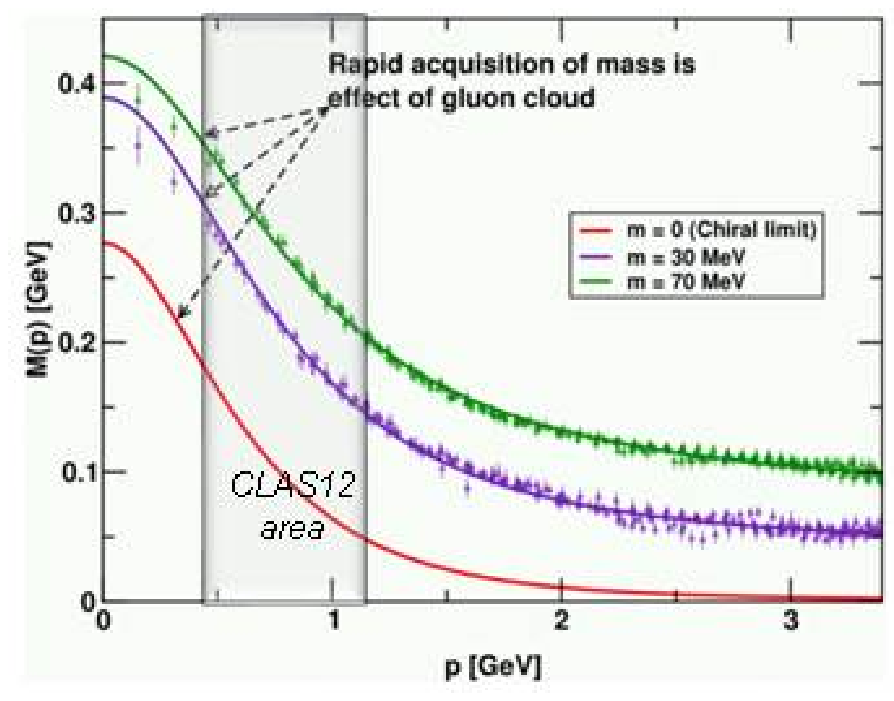} 
\caption{(Color Online) Left: Spectrum of $N^*$ states computed in LQCD starting from the QCD 
Lagrangian~\cite{Du12} will be mapped out with CLAS12 with the primary objective of the hybrid 
baryon search~\cite{Bu16,Dang16}. The contributions from the relevant quark and glue configurations 
to the resonance structure are shown by the horizontal bars connected by dashed lines to the boxes 
showing the $N^*$ masses and full widths with a pion mass of 524~MeV. The expected hybrid baryons with 
dominant contributions from glue are shown by the blue boxes. Right: Momentum dependencies of the 
running dressed quark mass from LQCD at two values of the bare quark masses, 30~MeV and 70~MeV,
are shown by points with error bars~\cite{Bo05} in comparison with the DSEQCD results (green and magenta 
lines) \cite{Cr14} for the same bare quark masses as in the LQCD studies. Evaluations in the chiral 
limit of massless bare quarks, which are close to the bare masses of the light $u$ and $d$ quarks, are 
currently available from DSEQCD only and shown by the red line~\cite{Cr14}. The dressed quark mass function 
in the momentum range that will be accessible for the first time from the studies of the $Q^2$ evolution of 
the resonance electrocouplings from exclusive meson electroproduction data with CLAS12~\cite{Ca16,Go16} 
is shown in the shadowed area, allowing us to explore the nature of $>$98\% of hadron mass, quark gluon 
confinement, and the emergence of the nucleon resonance structure from QCD \cite{Az13}.}
\label{hybqm}
\end{figure}

\section{Prospects for the future studies of excited nucleon states with the CLAS12}

After completion of the Jefferson Lab 12~GeV Upgrade Project, the CLAS12 detector in the upgraded 
Hall~B will be the only foreseen facility worldwide capable of studying nucleon resonances at the still 
unexplored ranges of the smallest photon virtualities 0.05~GeV$^2 < Q^2 < 0.5$~GeV$^2$ and the highest 
photon virtualities ever achieved in exclusive reaction measurements up to 12~GeV$^2$~\cite{Az13,Bu16,Go16}.

The studies of nucleon resonances at small photon virtualities are driven by the search for new 
states of baryon matter, the so-called hybrid-baryons~\cite{Dang16}. Small $Q^2$ is preferential for the 
observation of these new states that contain three dressed quarks and, in addition, glue as the structural 
component. The LQCD studies of the $N^*$ spectrum starting from the QCD Lagrangian~\cite{Du12} predict 
several such states of the positive parities shown in Fig.~\ref{hybqm} (left). The evaluations~\cite{Du12} 
were carried with much higher quark mass than needed to reproduce the pion mass. In order to estimate the 
physics masses of  hybrid states we corrected the results of \cite{Du12} reducing the predicted hybrid mass 
values by the differences between the experimental results on the masses of the known lightest $N^*$ of the 
same spin-parities as for the expected hybrid baryons and their values from LQCD. In the experiment with 
CLAS12 we will search for the hybrid signal as the presence of extra states in the conventional resonance 
spectrum of $J^P$=1/2$^+$, 3/2$^+$ in the mass range from 2.0~GeV to 2.5~GeV from the data of exclusive 
$KY$ and $\pi^+\pi^-p$ electroproduction off protons~\cite{Dang16}. The hybrid nature of the new baryon 
states will be identified looking for the specific $Q^2$ evolution of their electrocouplings. We expect the specific behavior of the hybrid state electrocouplings with $Q^2$ because one might imagine that the three quarks in a hybrid baryon should be in a color-octet state in order to create a colorless hadron in combination with the glue constituent in a color-octet state.  Instead, in regular baryons, constituent quarks should be in color-singlet 
state. So pronounced differences for quark configurations in the structure of conventional and hybrid baryons 
should results in a peculiar $Q^2$ evolution of hybrid baryon electrocouplings. The studies on the $N^*$ 
structure at low $Q^2$ over spectrum of all prominent resonances will also extend our knowledge on the
meson-baryon cloud in the resonance structure and will offer unique information on the $N^*$ excitations by 
longitudinal photons as $Q^2 \to 0$. The extension of amplitude analysis~\cite{Asz16,Sa16,Str16} and coupled 
channel analysis~\cite{Lee16,Ka16} methods, which were employed successfully in the studies of exclusive meson
photoproduction, for electroproduction off protons at small photon virtualities is of particular importance for 
the success of the aforementioned efforts.

The experiments on the studies of excited nucleon state structure in exclusive $\pi N$, $KY$, and 
$\pi^+\pi^-p$ electroproduction off protons at 5.0~GeV$^2 < Q^2 < 12.0$~GeV$^2$~\cite{Ca16,Go16a} 
are scheduled in the first year of running with the CLAS12 detector. For the first time electrocouplings 
of all prominent nucleon resonances will become available at the highest photon virtualities ever achieved 
in the exclusive electroproduction studies. These distance scales correspond to the still unexplored regime 
for the $N^*$ electroexcitation where the resonance structure is dominated by the quark core with almost 
negligible meson-baryon cloud contributions. The foreseen experiments offer almost direct access to the 
properties of dressed quarks inside $N^*$ states of different quantum numbers. Consistent 
results on the dressed quark mass function derived from independent analyses of the data on the
$\gamma_vpN^*$ electrocouplings of the resonances of distinctively different structure, such as radial 
excitations, spin-flavor flip, orbital excitations, will validate credible access to this fundamental 
ingredient of the non-perturbative strong interaction supported by the experimental data. The expected data 
on the $\gamma_vpN^*$ electrocouplings will provide for the first time access to the dressed quark mass 
function in the range of momenta/distance scales where the transition from the quark-gluon confinement to 
the pQCD regimes of strong interaction takes full effect, as is shown in Fig.~\ref{hybqm} (right). Exploring 
the dressed quark mass function at these distances will allow us to address the most challenging open problems 
of the Standard Model: on the nature of $>$98\% of hadron mass, quark-gluon confinement, and the emergence of 
the $N^*$ structure from the QCD Lagrangian~\cite{Cr15,Az13,Brod16,Du14}. 

In order to provide reliable evaluation of resonance electrocouplings at high photon virtualities, the 
reaction models for resonance electrocoupling extraction should be further developed, implementing explicitly 
quark degrees of freedom in the description of the non-resonant amplitudes. The efforts on implementation of 
the hand-bag diagrams for the part of non-resonant amplitudes in $\pi N$ electroproduction off protons in the 
$N^*$ region are underway~\cite{Kr16}.

\begin{acknowledgements}
This material is based upon work supported by the U.S. Department of Energy, Office of Science, Office of Nuclear Physics under contract DE-AC05-06OR23177.
\end{acknowledgements}



\end{document}